\documentclass[a4paper,12pt]{article}

\usepackage{amsmath,amssymb,amsfonts}% Typical maths resource packages
\usepackage{graphicx}% Packages to allow inclusion of graphics
\usepackage{color}% For creating coloured text and background
\makeatletter
\@addtoreset{equation}{section}
\renewcommand{\theequation}{\thesection.\@arabic\c@equation}
\makeatother
\usepackage{hyperref}% For creating hyperlinks in cross references
\usepackage{cite}
\definecolor{red}{rgb}{1,0,0}% Standard colours red, green, blue
\definecolor{green}{rgb}{0,1,0}
\definecolor{blue}{rgb}{0,0,1}
\definecolor{darkblue}{rgb}{0,0,0.5}
\definecolor{lightblue}{rgb}{.5,.5,1}
\definecolor{lightgray}{gray}{.87}% How you can define your own greys
\definecolor{Dark}{gray}{.20}
\definecolor{pink}{rgb}{.95,0.82,0.92}% How you can define your own colours
\definecolor{yellow}{rgb}{1,1,0}
\definecolor{lightyellow}{rgb}{1,1,.5}
\definecolor{purple}{rgb}{0.7,0,0.85}
\definecolor{darkgreen}{rgb}{0,0.5,0}
\definecolor{orange}{rgb}{0.8,0.2,0.2}
\def \be {\begin{equation}}
\def \ee {\end{equation}}
\def \bea {\begin{align}}
\def \eea {\end{align}}

\def \rr {\raise.35ex\hbox{\small $\prime$}\kern-.17em{\mbox{\large $\imath$}}}
\def \del {\partial}
\def \dels {\partial\kern-.5em / \kern.5em}
\def \As {{A\kern-.5em / \kern.5em}}
\def \Ds {D\kern-.7em / \kern.5em}

\def \eps {\epsilon}

\newcommand{\detail}[1]{}

\newcommand{\hide}[1]{}

\newcommand{\explanation}[1]{}

\pdfoutput=1

\begin{document}

\pagestyle{plain}

\begin{titlepage}
\begin{flushright}
\vspace*{-24pt}
\small\tt
OU-HET 1008
\vspace{24pt}
\end{flushright}

\begin{center}

\noindent
\textbf{\LARGE
%%% 03/12
Vacuum Energy at Apparent Horizon \\
in Conventional Model of Black Holes
%On the Near-Horizon Geometry \\
%of an Evaporating Black Hole 2: \\
%Collapsing Spherical Null Shells \\
\vskip.6cm
}
\vskip .5in
\renewcommand{\thefootnote}{\fnsymbol{footnote}}
{\large 
Pei-Ming Ho%
${}^{a}$%
\footnote[2]{e-mail: pmho@phys.ntu.edu.tw},
%Hikaru~Kawai${}^b$
%\footnote[2]{e-mail: hkawai@gauge.scphys.kyoto-u.ac.jp},
Yoshinori Matsuo%
${}^{a,b}$%
\footnote[3]{e-mail: matsuo@phys.ntu.edu.tw},
Shu-Jung Yang%
${}^{a}$%
\footnote[4]{e-mail: dodolong0619@gmail.com}
}
\\
\vskip 10mm
\renewcommand{\thefootnote}{\arabic{footnote}}
{\sl 
${}^{a}$
Department of Physics and Center for Theoretical Physics, \\
National Taiwan University, Taipei 106, Taiwan,
R.O.C. 
\\
\vskip 3mm
${}^{b}$
Department of Physics, Osaka University, \\
Toyonaka, Osaka 560-0043, Japan
}

\vskip 3mm
\vspace{60pt}
\begin{abstract}

%5/22+
For a black hole of Schwarzschild radius $a$,
we argue that the back-reaction of the vacuum energy-momentum tensor
is in general important at the apparent horizon when
the time scale of a process is larger than order $a$.
In particular,
in a double-shell model,
we show that the ignorance of the back-reaction
%of the ingoing vacuum energy flux
leads to a divergence in the outgoing vacuum energy flux.
The main result of this paper is that,
once the back-reaction is included,
the vacuum energy density for observers on top 
of the trapping horizon in vacuum is given by
${\cal E} \simeq -1/2\ell_p^2 a^2$,
which is of the same order of magnitude
as the classical matter but opposite in sign.
Remarkably, this formula is independent of
both the details about the collapsing matter
and the vacuum energy-momentum tensor.
%5/22-

\end{abstract}
\end{center}

\end{titlepage}

\baselineskip 18pt

\noindent\rule{\textwidth}{1pt}

\tableofcontents

\vskip 12pt

\noindent\rule{\textwidth}{1pt}

\setcounter{page}{1}
\setcounter{footnote}{0}
\setcounter{section}{0}

\section{Introduction}

In the conventional model of black holes,
it is often assumed that the back-reaction of
the vacuum energy-momentum tensor $\langle T_{\mu\nu} \rangle$
can be ignored at the horizon,
and the static Schwarzschild metric provides a good approximation.
A necessary condition for the consistency of this assumption is that
the vacuum energy-momentum tensor is finite
in generic free-falling frames.
In terms of the light-cone coordinates $(u, v)$ defined by
the asymptotic Minkowski space at large distance,
this demands that
$\langle T_{uu} \rangle$ and $\langle T_{uv} \rangle$
are very small at the horizon.
One must therefore have a negative ingoing energy flux
$\langle T_{vv} \rangle$ of order $\mathcal{O}(1/a^4)$
for a black hole of the Schwarzschild radius $a$
to account for the evaporation.
Therefore,
In the conventional model,
the ingoing negative vacuum energy dominates at the trapping horizon.

We argue that the back-reaction of 
the ingoing negative vacuum energy $\langle T_{vv} \rangle$
is important when the time scale of a process
is well above order $\mathcal{O}(a)$.
As a concrete example,
we consider a collapsing thin shell followed by 
a second shell of arbitrarily small mass,
as a small perturbation to the first shell.
There would be a moment when 
the outgoing vacuum energy flux diverges in a generic free-falling frame
if the first shell falls under its Schwarzschild radius
before the second shell crosses the Schwarzschild radius for the whole system.
We show that this divergence is removed by taking into consideration
the back-reaction of the ingoing negative vacuum energy $\langle T_{vv} \rangle$.

With the ingoing negative vacuum energy included 
in the semi-classical Einstein equation
\be
G_{\mu\nu} = \ell_p^2 \langle T_{\mu\nu}\rangle,
\label{scEE}
\ee
where $\ell_p^2 \equiv 8\pi G_N$
($G_N$ is the Newton constant),
we study the dynamical geometry of a small neighborhood of the trapping horizon
during the gravitational collapse.
We use the convention that $\hbar = c = 1$
so that $\ell_p$ can be identified with the Planck length.

The trapping horizon 
(the world-history of the apparent horizon)
is time-like in vacuum due to the negative vacuum energy flux.
The main result of this paper is that,
for observers on top of the trapping horizon,
the vacuum energy density is given by a universal formula:
${\cal E} \simeq -1/2\ell_p^2 a^2$ \eqref{E}.
This expression is independent of the details of the vacuum energy-momentum tensor,
as long as it is dominated by the ingoing negative vacuum energy at the horizon.
It is also independent of the collapsing matter and the collapsing process,
as long as the trapping horizon exists.
Note that this gauge-invariant quantity ${\cal E}$,
as it is inversely proportional to $\ell_p^2$,
is of the same order as the classical mass density of the black hole
but negative in value.

The energy density ${\cal E}$ corresponds to a negative energy flux
of power $P = - 2\pi/\ell_p^2$ \eqref{P} falling through the apparent horizon
at the speed of light.
At this order of approximation,
this is the only reason for the decrease in the black hole mass over time.
Eventually,
the total negative energy behind the apparent horizon
cancels the energy of the collapsed matter.
The matter under the apparent horizon is not really evaporated
but coexists with an equal magnitude of negative vacuum energy.
The holographic principle is not expected to hold in this model
since it admits a macroscopic amount of negative energy.
On the other hand,
the existence of a gauge-invariant quantity that becomes large at the horizon
casts doubt on the reliability of the conventional model,
as such gauge-invariant operators might appear in the Lagrangian
of certain effective theories.

The plan of this paper is as follows.
We review in Sec.\ref{DFU}
the widely applied model of vacuum energy-momentum tensor
proposed by Davies, Fulling and Unruh \cite{Davies:1976hi,Davies:1976ei},
as a concrete example of the conventional model of black holes.
In Sec.\ref{ValidityConstantBackground},
we argue that the back-reaction fo the vacuum energy-momentum tensor
is important when the time scale of a process
is of order $\mathcal{O}(a)$ or larger
for a black hole of Schwarzschild radius $a$.
We demonstrate in a concrete model of double shells in Sec.\ref{DoubleShell}
that if the back-reaction of the ingoing vacuum energy flux $\langle T_{vv}\rangle$
is not properly taken into consideration,
there would be a divergence in the outgoing vacuum energy flux $\langle T_{uu} \rangle$.
In Sec.\ref{Generic}, 
including the back-reaction of $\langle T_{vv}\rangle$ 
in the ingoing Vaidya metric as
a solution to the semi-classical Einstein equation,
we compute the vacuum energy density for observers
staying on top of the trapping horizon
and found the universal formula \eqref{E},
without assuming an explicit expression of the vacuum energy-momentum tensor.
Finally, 
we comment in Sec.\ref{conclusion}
the implications of our results
and compare our results for the conventional model
with other models of black holes.

\section{Review of Conventional Model}
\label{DFU}

In this work,
we focus on 4D spacetime with spherical symmetry.
In general,
the metric can be written as
\be
ds^2 = - C(u, v) du dv + r^2(u, v) d\Omega^2,
\label{metric}
\ee
where $u$, $v$ are the light-cone coordinates,
and $d\Omega^2 = d\theta^2 + \sin^2\theta d\phi^2$
is the metric of the 2-sphere. 
The sphere at $r(u,v) = r$ has the area of $4\pi r^2$.

We assume in this section and Sec.\ref{DoubleShell}
that the vacuum expectation value of the energy-momentum tensor
$\langle T_{\mu\nu} \rangle$
is given by the model proposed by
Davies, Fulling and Unruh \cite{Davies:1976hi,Davies:1976ei},
that is,
\begin{align}
\langle T_{uu} \rangle &= 
- \frac{1}{12\pi r^2} C^{1/2}\del_u^2 C^{-1/2}
+ \frac{1}{16\pi r^2}\{U, u\},
\label{Tuu-4D}
\\
\langle T_{vv} \rangle &=
- \frac{1}{12\pi r^2} C^{1/2}\del_v^2 C^{-1/2}
+ \frac{1}{16\pi r^2}\{V, v\},
\label{Tvv-4D}
\\
\langle T_{uv} \rangle &=
\frac{1}{12\pi r^2 C^3}\left(C\del_u\del_v C - \del_u C\del_v C\right),
\label{Tuv-4D}
\\
\langle T_{\theta\theta} \rangle &= \langle T_{\phi\phi} \rangle = 0,
\label{Tthth-4D}
\end{align}
where $U$ and $V$ are the light-cone coordinates used 
to specify the vacuum state \cite{Davies:1976hi},
and $\{\cdot, \cdot\}$ is the Schwarzian derivative defined by
\begin{align}
\{ f, u \} \equiv \left(\frac{\frac{d^2 f}{du^2}}{\frac{df}{du}}\right)^2
- \frac{2}{3} \frac{\left(\frac{d^3 f}{du^3}\right)}{\left(\frac{df}{du}\right)}.
\label{Schwarzian}
\end{align}
We refer to the resulting semi-classical theory of black holes
as the DFU model.
It is a concrete representative of the conventional model of black holes.

As an example,
consider a collapsing thin shell of areal radius $R_0(u)$.
The space inside the shell $(r < R_0(u))$
remains in the Minkowski vacuum.
The Minkowski metric is
\be
ds^2 = - dU dV + r^2(U, V) d\Omega^2,
\ee
in terms of the light-cone coordinates $U$, $V$,
where
\begin{align}
r(U, V) \equiv \frac{V - U}{2}.
\label{taur-UV}
\end{align}

When the back-reaction of the vacuum energy-momentum tensor is ignored,
the space outside the thin shell $(r \geq R_0(u))$
is given by the Schwarzschild metric
\be
ds^2 = - \left(1-\frac{a_0}{r(u, v)}\right)du dv + r^2(u, v) d\Omega^2,
\label{Schwarzschild-metric}
\ee
where $a_0$ is proportional to the thin shell's mass.
and $r(u, v)$ satisfies
\be
\frac{\del r}{\del u} = - \frac{\del r}{\del v}
= - \frac{1}{2}\left(1 - \frac{a_0}{r}\right).
\label{drdu}
\ee

On the trajectory of the collapsing shell at $r = R_0(U)$,
we have $V = U + 2R_0(U)$
according to eq.\eqref{taur-UV}.
The continuity of the metric
across the shell implies that
%\begin{align}
%& \left. dU dV\right|_{r=R_0(U)} 
%= \left.\left(1-\frac{a_0}{R_0(U)}\right) du dv\right|_{r=R_0(U)},
%\end{align}
%which can be solved by
\begin{align}
\frac{dU}{du}
%&=
%\frac{1}{1+2\frac{dR_0}{dU}}\left[
%\frac{dR_0}{dU}
%+\sqrt{
%\left(\frac{dR_0}{dU}\right)^2
%+\left(1-\frac{a_0}{R_0}\right)\left(1+2\frac{dR_0}{dU}\right)
%}
%\right]
%\nn \\
%\label{dUdu-thinshell}
%\end{align}
%(The other solution is discarded because we need $dU/du > 0$.)
%\begin{align}
%\frac{dU}{du}
\simeq 
- \frac{R_0-a_0}{2a_0\frac{dR_0}{dU}}
+ \frac{\left[1+2\frac{dR_0}{dU}
+4\left(\frac{dR_0}{dU}\right)^2\right]}{8a_0^2\left(\frac{dR_0}{dU}\right)^3}(R_0-a_0)^2
+ \mathcal{O}\left((R_0-a_0)^3\right),
\label{dUdu-1}
\end{align}
as an expansion of $(R_0 - a_0)$.
For finite non-vanishing $dR_0/dU$,
this reproduces the conventional result
for Hawking radiation at large $r$
as $R_0 \rightarrow a_0$
\cite{Davies:1976ei}:
\begin{align}
\frac{1}{16\pi r^2}\{U, u\}
= \frac{1}{16\pi r^2}\frac{1}{12a_0^2}
+ \mathcal{O}\left((R_0-a_0)^2\right).
\label{Tuu-0}
\end{align}

Using eqs.\eqref{Tuu-4D}--\eqref{Tuv-4D},
one can compute the vacuum energy-momentum tensor 
at the moment of crossing $R_0 = a_0$
at the lowest order \cite{Davies:1976ei}.
In the limit $r \rightarrow a_0$,
it is
\begin{align}
\langle T_{uu} \rangle &=
\frac{1}{24\pi r^2} \left(\frac{3a_0^2}{8r^4} 
- \frac{a_0}{2r^3} + \frac{1}{8a_0^2}\right)
\longrightarrow \mathcal{O}((r-a_0)^2),
\label{T2uu}
\\
\langle T_{vv} \rangle &=
\frac{1}{24\pi r^2} \left(\frac{3a_0^2}{8r^4} - \frac{a_0}{2r^3}\right)
\longrightarrow - \frac{1}{192\pi a_0^4} + \mathcal{O}((r-a_0)),
\label{T2vv}
\\
\langle T_{uv} \rangle &=
\frac{1}{24\pi r^2} \left(\frac{a_0^2}{2r^4} - \frac{a_0}{2r^3}\right)
\longrightarrow \mathcal{O}((r-a_0)).
\label{T2uv}
\end{align}
One can check that the regularity conditions 
\cite{Christensen:1977jc,Fulling:1977jm}:
\begin{align}
(r-a_0)^{-2}|\langle T_{uu} \rangle|
&< \infty,
\label{reg-Tuu}
\\
(r-a_0)^{-1}|\langle T_{uv} \rangle|
&< \infty,
\label{reg-Tuv}
\\
|\langle T_{vv} \rangle| 
&< \infty
\label{reg-Tvv}
\end{align}
are satisfied.
Notice that the vacuum energy-momentum tensor
is dominated by an ingoing negative energy flux $\langle T_{vv} \rangle$
\eqref{T2vv} on the horizon.
%For the regularity of the energy-momentum tensor,
%it is crucial that the 0-th order term is
%$\frac{1}{16\pi}\frac{1}{12a_0^2}$
%and the first order term in $(R_0-a_0)$ is absent.
%5/22-

\section{Range of Validity of Constant Background}
\label{ValidityConstantBackground}

In the previous section,
we reviewed the single-shell model of black holes.
The calculation of the vacuum energy-momentum tensor
appears to be self-consistent without taking into account its back-reaction.
However, 
this does not guarantee the same level of consistency
for more realistic models.
In this section,
we examine the time scale over which
the constant background approximation 
(without back-reaction)
is at the risk of breaking down.
In the next section,
we will see that the ignorance of back-reaction
does lead to inconsistency in more general cases.

With the back-reaction ignored,
the time-independent Schwarzschild metric \eqref{Schwarzschild-metric}
can be written as
\be
ds^2 = - \left(1-\frac{a_0}{r}\right) du^2
- 2dudr + r^2 d\Omega^2,
\ee
where $u$ is an outgoing light-cone coordinate.
For simplicity,
we assume that the surface of the shell collapses at the speed of light,
i.e.
\be
\frac{dR_0}{du} = - \frac{1}{2}\left(1-\frac{a_0}{R_0}\right),
\label{R-eq}
\ee
where $R_0(u)$ is the areal radius of the collapsing shell.

Suppose that,
at a certain moment $u = u_1$,
\be
R_0(u_1) - a_0 = L \ll a_0,
\ee
eq.\eqref{R-eq} can be approximately solved by
\be
R_0(u) \simeq a_0 + L e^{-\frac{u-u_1}{2a_0}}
\ee
for $u > u_1$.
When the shell is separated from the Schwarzschild radius
only by a small distance of the order of the Planck length,
\be
R_0(u_2) - a_0 \sim \frac{\ell_p^2}{a_0},
\label{Ru2-a}
\ee
the time $u_2$ is given by
\be
\Delta u \equiv u_2 - u_1 \sim 2a_0 \log\left(\frac{La}{\ell_p^2}\right) \gg 2a_0,
\label{u2-u1}
\ee
as long as $L$ is not too small.

Since the Schwarzschild radius $a_0$ is assumed to be a constant in this calculation,
eq.\eqref{Ru2-a} is invalid if 
the change in the Schwarzschild radius
$\Delta a_0 \equiv a_0(u_2) - a_0(u_1)$
is larger than $\ell_p^2/a_0$.
In fact,
according to the conventional formula for Hawking radiation
\be
\frac{da_0}{du} \sim - \frac{\ell_p^2}{a_0^2},
\label{dadu}
\ee
we have
\be
\Delta a_0 \simeq \left|\frac{da_0}{du}\right|(u_2-u_1)
\sim \frac{\ell_p^2}{a_0^2} \left[2a_0\log\left(\frac{La}{\ell_p^2}\right)\right]
\gg \frac{\ell_p^2}{a_0}.
\label{Deltaa}
\ee
Hence eq. \eqref{Ru2-a} is a terrible estimate
of the difference $(R_0-a_0)$.
In general,
back-reaction is important when
the time scale $\Delta u$ of the physical process under consideration is
\be
\Delta u \gtrsim \mathcal{O}(a_0),
\ee
at least for quantities sensitive to $(R_0-a_0)$
when it is of the order of $\ell_p^2/a_0$ or smaller.
We will demonstrate in Sec.\ref{DoubleShell}
that the negligence of the back reaction may lead to
a divergence in $\langle T_{uu} \rangle$.

Incidentally,
the study of general static black-hole solutions to the semi-classical Einstein equation \eqref{scEE}
\cite{Ho:2018fwq}
also shows that the Schwarzschild metric is in general
not necessarily a good approximation for
\be
r - a_0 \lesssim \mathcal{O}\left(\frac{\ell_p^2}{a_0}\right).
\label{r-a<k}
\ee
The heuristic reason is the following.
Naively,
the semi-classical Einstein equation \eqref{scEE}
is simplified to the vacuum equation $G_{\mu\nu} = 0$ 
at the 0-th order of the $\hbar$-expansion
(which is equivalent to the $\ell_p^2$-expansion in vacuum).
However,
as the Schwarzschild metric involves the factor $\left(1-\frac{a_0}{r}\right)$,
which introduces a factor of $\ell_p^2/a_0^2$
when $r - a_0$ is of order $\mathcal{O}(\ell_p^2/a_0)$,
a factor of $\ell_p^2$ is introduced on the left hand side
of the semi-classical Einstein equation,
so that it is no longer obviously consistent to ignore the right hand side 
in the limit $\ell_p^2 \rightarrow 0$.

\section{Double-Shell Model}
\label{DoubleShell}

As a process with a longer time scale than the single shell model,
we consider the scenario involving two thin shells.
We will show that,
as a result of ignoring the back-reaction of the ingoing vacuum energy flux \eqref{T2vv},
the standard calculation leads to a divergence in the outgoing vacuum energy flux \eqref{Tuu-4D}.

Assume that the two shin shells are of masses $m_1$, $m_2$
and radii $R_1$, $R_2$ ($R_1 < R_2$).
The metric is 
\be
ds^2 = \left\{
\begin{array}{ll}
- dUdV + r^2 d\Omega^2 &
\mbox{inside the first shell},
\\
- \left(1-\frac{a_1}{r}\right) du_1 dv_1 + r^2 d\Omega^2 & 
\mbox{between the two shells},
\\
- \left(1-\frac{a_2}{r}\right) du dv + r^2 d\Omega^2 &
\mbox{outside the second shell},
\end{array}
\right.
\label{double-shell-metric}
\ee
where we used the coordinates $(U, V)$ for the Minkowski space
inside the first (inner) shell,
$(u_1, v_1)$ for the Schwarzschild metric between the shells,
and $(u, v)$ for the Schwarzschild metric outside the 2nd (outer) shell.
The Schwarzschild radii are
\begin{align}
a_1 = 2G_N m_1,
\qquad
a_2 = 2G_N (m_1 + m_2).
\end{align}

Assuming for simplicity
that both shells are falling at the speed of light,
the continuity conditions across the thin shells
determine the ratios of increments in different time coordinates:
\begin{align}
\frac{du_2}{du_1} = \frac{R_2-a_1}{R_2-a_2},
\qquad
\frac{du}{dU}
%= \frac{du_2}{du_1}\frac{du_1}{dU}
= \left(\frac{R_2-a_1}{R_2-a_2}\right)\left(\frac{R_1}{R_1-a_1}\right).
\end{align}

At a moment when both shells are close to their Schwarzschild radii,
let
\begin{align}
R_1 = a_1 + \eps_1,
\qquad
R_2 = a_2 + \eps_2 = a_1 + \eps,
\end{align}
where $\eps_1$ and $\eps_2$ are small,
and
\be
\eps \equiv 2m_2 + \eps_2.
\ee
The outgoing vacuum energy flux in the limit $\eps_2 \rightarrow 0$ is given by
\begin{align}
\langle T_{uu} \rangle 
= \left\{
- \frac{1}{32\pi a_2^4}
\left[\left(\frac{\eps_1}{\eps}\right)^2 - 1\right]
+ \mathcal{O}\left(\frac{\eps_1}{\eps}\right)
\right\}
\frac{\eps_2^2}{a_2^2} + \mathcal{O}(\eps_2^3),
\label{check-Tuu}
\end{align}
so that,
as long as both $\eps$ and $\eps_1$ are finite,
the regularity condition \eqref{reg-Tuu} is satisfied.

Let us now consider the following situation.
The first shell of mass $m_1$ has already collapsed into
its Schwarzschild radius $a_1$ so that $R_1 < a_1$.
We take the second shell to have an infinitesimal mass $m_2 \rightarrow 0$
so that this system is in practice
indistinguishable from a single shell of mass $m_1$.
However, 
as the 2nd shell approaches $a_1$
(so that $R_2 \sim a_2 \simeq a_1 > R_1$),
there are moments when
\be
|\eps_1| \gg |\eps|,
\label{eps1ggeps}
\ee
as $|\eps| \equiv |R_2-a_1|$ can be arbitrarily small,
so that the outgoing energy flux \eqref{check-Tuu}
is arbitrarily large in a free-falling frame!
In fact, 
$\langle T_{uu} \rangle$ diverges when $R_2 = a_1$.

There are two ways to interpret this result.
One may say that the reason for the divergence in $\langle T_{uu} \rangle$
is that we should have used a time-dependent metric,
rather than the constant Schwarzschild metrics \eqref{double-shell-metric},
to properly describe the black-hole evaporation.
The back-reaction of the vacuum energy-momentum tensor should not be ignored.

\begin{figure}[h]
\vskip0em
\center
\includegraphics[scale=0.45,bb=0 80 450 320]{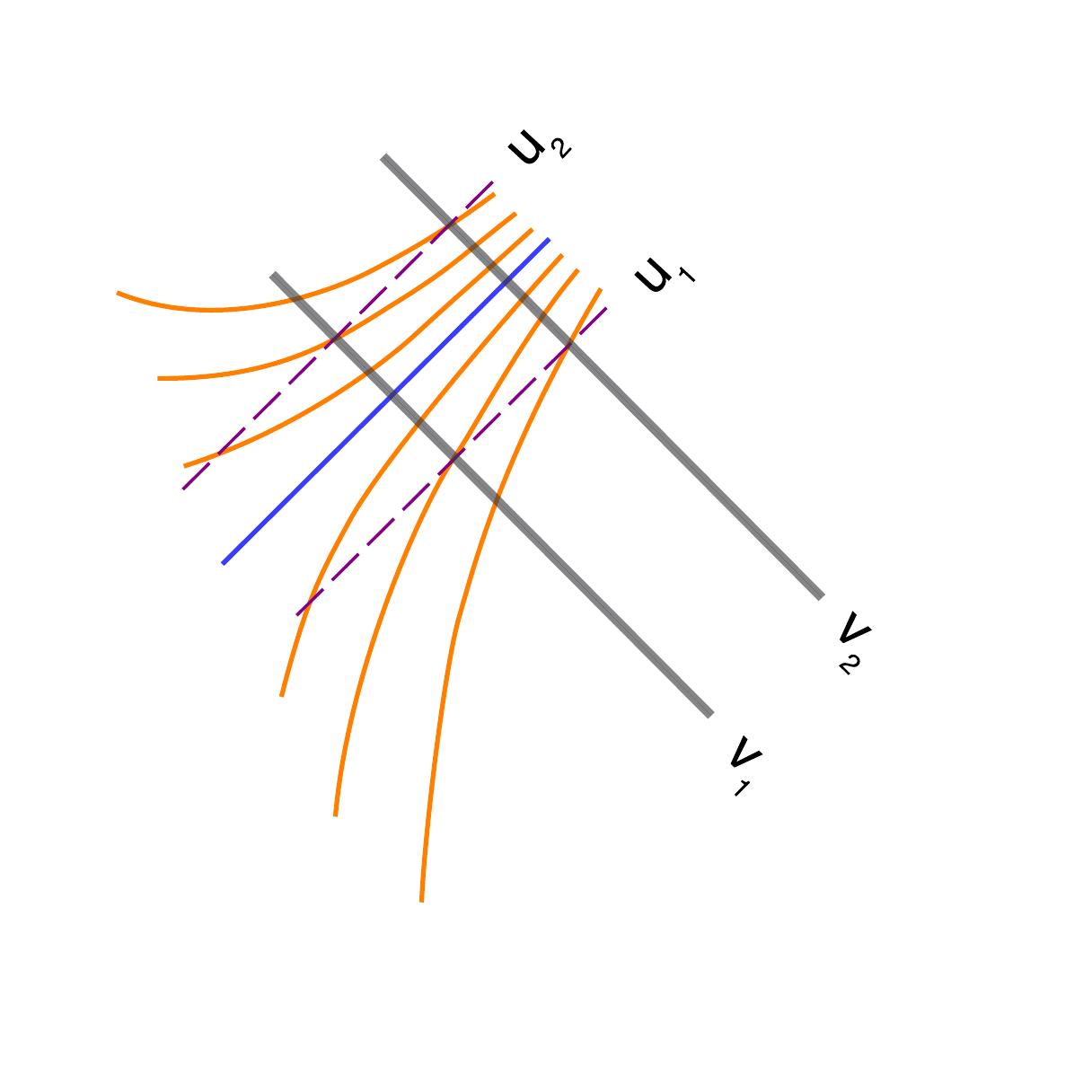}
\includegraphics[scale=0.45,bb=0 80 300 290]{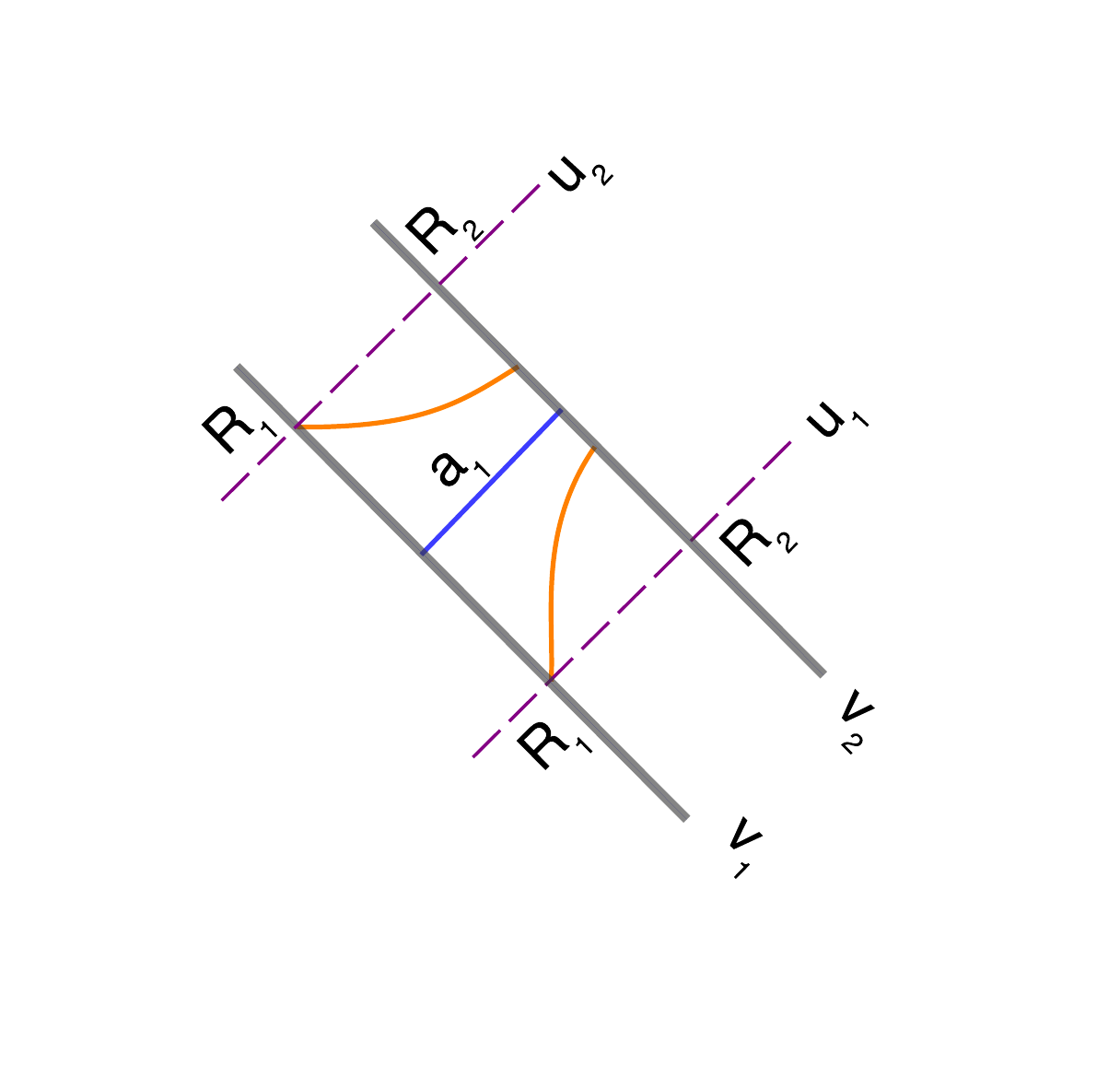}
\vskip1em
(a)\hskip17em(b)
\caption{\small
(a) A small neighborhood of the future horizon in a Penrose diagram:
The orange curves are constant-$r$ curves.
The null shell at $v_2$ has a larger (smaller) areal radius
at $u = u_1$ ($u = u_2$)
in comparison with the null shell at $v_1$.
(b) For a constant Schwarzschild radius,
either we have $R_2 > R_1 > a_1$ or $a_1 > R_1 > R_2$.
Hence we always have $|\eps_1/\eps| < 1$.
}
\label{Classical-Horizon}
\vskip0em
\end{figure}

Alternatively,
one may also say that,
with the back-reaction ignored,
the configuration considered above 
can never occur if both shells are initially outside the Schwarzschild radius.
During the formation process,
the geometry is given by the static Schwarzschild metric
with constant Schwarzschild radius $a_2$ outside the outer shell
and constant Schwarzschild radius $a_1$ between the two shells.
One can then see that 
two shells cross the horizon at $r = a_1$ at the same retarded time $u$. 
At arbitrary retarded time,
we always have $|\eps| \geq |\eps_1|$,
regardless of whether the shells are inside or outside the horizons.
It is impossible to have $|\eps| \ll |\eps_1|$,
and there would be no divergence in the outgoing energy flux.
This is demonstrated pictorially in Fig.\ref{Classical-Horizon}.
The crucial point is that,
for a constant Schwarzschild background,
the two shells must be both under or above the Schwarzschild radius $a_1$
at any instant of $u$.

However, 
in the conventional model of black holes,
one should be allowed to consider some configurations in which
a part of the collapsing matter is already under the horizon,
while the rest of the collapsing matter is still outside the horizon.
A situation similar to $|\eps_1| \gg |\eps| \sim 0$
in our double-shell configuration would typically occur.
We would then get diverging outgoing energy flux in a local free-falling frame 
if we ignore the back-reaction.

The conclusion is thus that
the conventional model of black holes 
is incapable of properly describing such a configuration
unless the back-reaction of the vacuum energy-momentum tensor
should be taken into account.

On the other hand,
it is possible to find a consistent description of the black-hole geometry
by taking into consideration
only the back-reaction of the ingoing vacuum energy flux $\langle T_{vv}\rangle$.
We can use thin shells of tiny negative mass
to represent the ingoing negative vacuum energy,
so that the Schwarzschild radius outside 
a thin shell of negative mass is slightly smaller than the one inside,
and the decrease in the black hole mass can be properly described.

\begin{figure}[h]
\vskip0em
\center
\includegraphics[scale=0.45,bb=0 100 500 350]{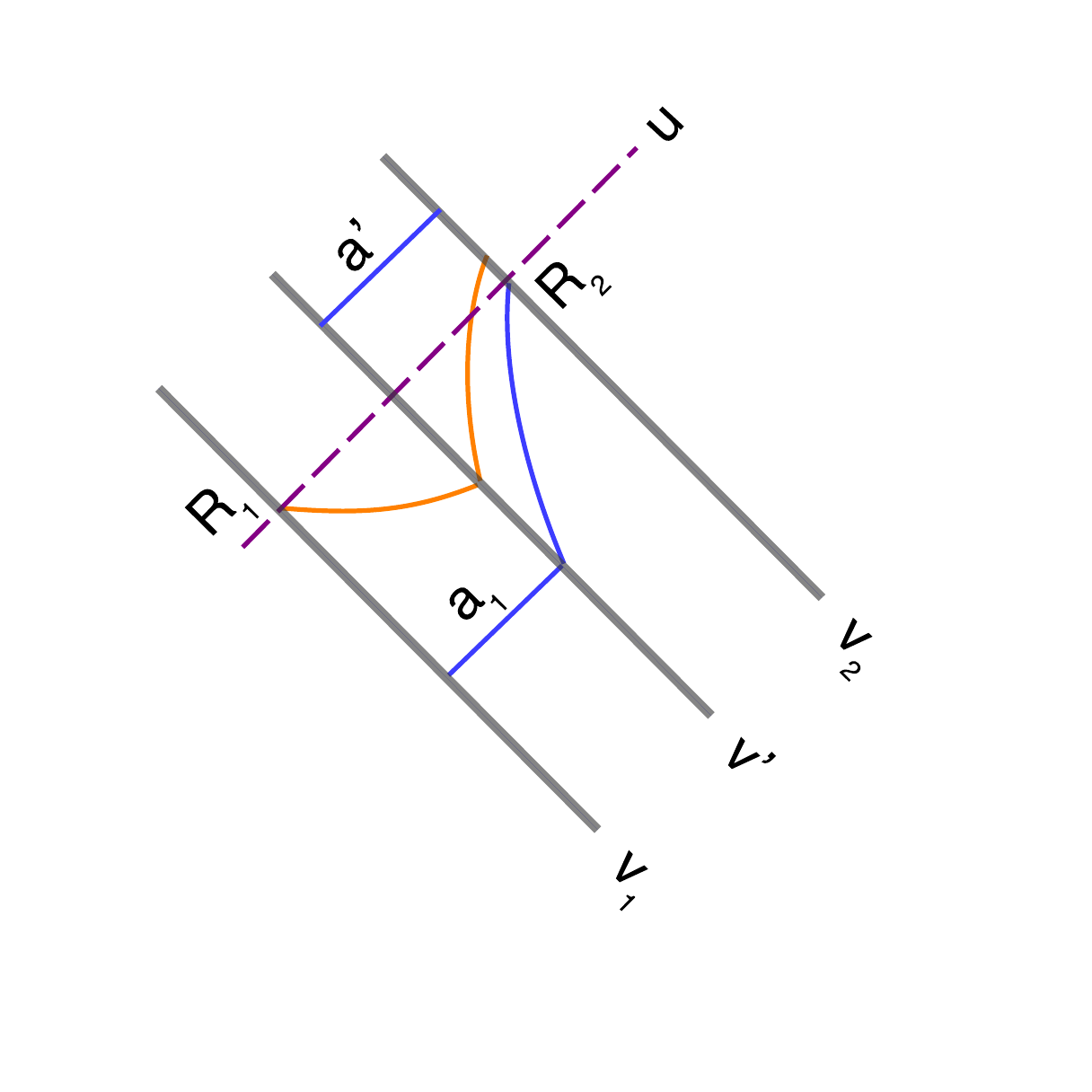}
\vskip2em
\caption{\small
Part of the Penrose diagram with two null shells at $v_1$ and $v_2$,
with a negative shell in the middle at $v'$.
The Schwarzschild radius is $a_1$ for $v \in (v_1, v')$,
and it is $a'$ for $v \in (v', v_2)$.
The orange curve is the constant-$r$ curve for $r = R_1$.
The blue curve is the constant-$r$ curve for $r = a_1$,
which is taken to be very close to $R_2$ at the instant $u$.
At the instant $u$,
we have $|\eps|$ arbitrarily close to $0$,
so $|\eps_1|/|\eps|$ is arbitrarily large as we considered in eq.\eqref{eps1ggeps}.
}
\label{Discrete-Horizon}
\vskip1em
\end{figure}

For instance, 
for the configuration considered above,
we introduce a thin shell with a negative mass between 
the two shells of collapsing matter
to describe approximately the negative vacuum energy between the shells. 
See Fig.\ref{Discrete-Horizon}.
In this case, the distance between $a_1$ and 
the radius at the negative mass shell cannot be smaller than $|\epsilon_1|$ 
for the same reason depicted in Fig.\ref{Classical-Horizon}. 
There is no divergence at the negative mass shell. 
For the second shell of collapsing matter,
$|\epsilon|$ can be smaller than $|\epsilon_1|$
however,
the outgoing energy flux no longer diverges in a local free-falling frame
due to the back-reaction of the shell of negative mass at $v = v'$.

The lesson we learned from this exercise is that 
while the ignorance of back-reaction
of the vacuum energy-momentum tensor is not good for 
generic configurations,
it is possible to find a good approximate description
of the near-horizon geometry
by including the ingoing negative vacuum energy flux alone.

In Sec.\ref{Generic},
we will present an approximate solution around the trapping horizon to
the semi-classical Einstein equation
in which the back-reaction of the ingoing vacuum energy flux
$\langle T_{vv}\rangle$ is incorporated,
although the other two components
$\langle T_{uu}\rangle$ and $\langle T_{uv}\rangle$
of the vacuum energy-momentum tensor are still ignored.
The vacuum energy-momentum tensor
is of course not exactly the same
as a continuous ingoing negative energy flux from past infinity,
yet this may still provide a good approximation of the geometry
in a sufficiently small neighborhood around the trapping horizon.
The geometry far away from the trapping horizon is
still expected to be well approximated by 
a time-dependent Schwarzschild background.

\section{Generic Null Shell}
\label{Generic}

Recall that both $\langle T_{uu} \rangle$ and $\langle T_{uv} \rangle$
must vanish at the horizon as consistency conditions \eqref{reg-Tuu}, \eqref{reg-Tuv}
for the static Schwarzschild background to be a valid 0-th order approximation,
In the following, 
we will assume that $\langle T_{uu} \rangle$ and $\langle T_{uv} \rangle$ are negligible
in the semi-classical Einstein equation
for a small neighborhood of the trapping horizon,
but we will not assume that the vacuum energy-momentum tensor $\langle T_{\mu\nu} \rangle$
is given by any specific expression.

Setting $\langle T_{uu} \rangle = \langle T_{uv} \rangle = 0$
for the background,
$\langle T_{vv} \rangle$ is a conserved ingoing energy flux,
so we have the ingoing Vaidya metric
\be
ds^2 = - \left(1-\frac{a(v)}{r}\right)dv^2 + 2dvdr + r^2 d\Omega^2
\label{ingoing-Vaidya-metric}
\ee
for a generic, spherically symmetric collapsing sphere.
This should be a good approximation of a sufficiently small neighborhood
of the trapping horizon.

The ingoing energy flux
\be
T_{vv} = \frac{1}{\ell_p^2}\frac{a'(v)}{r^2}
\label{Tvv-a}
\ee
includes both the collapsing matter of positive energy
and the negative vacuum energy flux.
Let the surface of the collapsing shell be located at $v = v^*$.
We have positive energy for the collapsing shell 
and negative energy for the vacuum:
\begin{align}
a'(v) \geq 0 & \quad \mbox{for} \quad v < v^*,
\\
a'(v) \leq 0 & \quad \mbox{for} \quad v > v^*.
\end{align}

The ingoing Vaidya metric can also be expressed 
in terms of the light-like coordinates $(u, v)$ as
\be
ds^2 = - C(u, v) dudv + R^2(u, v) d\Omega^2,
\label{ds2}
\ee
where
$C(u, v)$ and $R(u, v)$ satisfy
\begin{align}
\frac{\del R(u, v)}{\del u} &=
- \frac{1}{2} C(u, v),
\label{dRuvdu-null}
\\
\frac{\del R(u, v)}{\del v} &=
\frac{1}{2}\left(1-\frac{a(v)}{R(u, v)}\right).
\label{dRuvdv-null}
\end{align}
\noindent

The consistency of 
eqs.\eqref{dRuvdu-null} and \eqref{dRuvdv-null}
demands that
\be
\frac{\del}{\del v}C(u, v) = \frac{a(v)}{2 r^2(u, v)}C(u, v),
\ee
which can be solved by
\be
C(u, v') = C(u, v) e^{\int_{v}^{v'} dv'' \;\frac{a(v'')}{2r^2(u,v'')}}.
\ee 
For the asymptotic Minkowski space at large $v$,
\be
C(u, \infty) = 1,
\ee
we find
\be
C(u, v) = e^{- \int_{v}^{\infty} dv' \;\frac{a(v')}{2r^2(u,v')}} > 0.
\label{C-sol}
\ee

\subsection{Around Trapping Horizon}

In this section, 
we derive the main result of the paper eq.\eqref{E},
which is a universal formula of the energy density
for observers staying on top of the trapping horizon.
The trapping horizon is sometimes considered as
the geometric feature that characterizes a black hole \cite{Hayward:2005gi}.
Since we have assumed spherical symmetry,
it is convenient to define the trapping horizon by
the foliation of the space-time into symmetric 2-spheres.

Recall that
a symmetric 2-sphere $S$ is a trapped surface
if both ingoing and outgoing null geodesics orthogonal to $S$
have negative expansion.
This means that
$\del_u R < 0$ and $\del_v R < 0$
in terms of the areal radius $R(u, v)$
as a function of $u$ and $v$.
(In contrast,
$\del_u R < 0$ and $\del_v R > 0$
for the Minkowski space-time.)
The boundary of a trapped region ---
a 3D region composed of trapped surfaces ---
is called a trapping horizon,
where $\del_v R = 0$.
A 2D space-like section of the trapping horizon 
is an apparent horizon.

\begin{figure}[h]
\vskip0em
\center
\includegraphics[scale=0.5,bb=0 10 450 400]{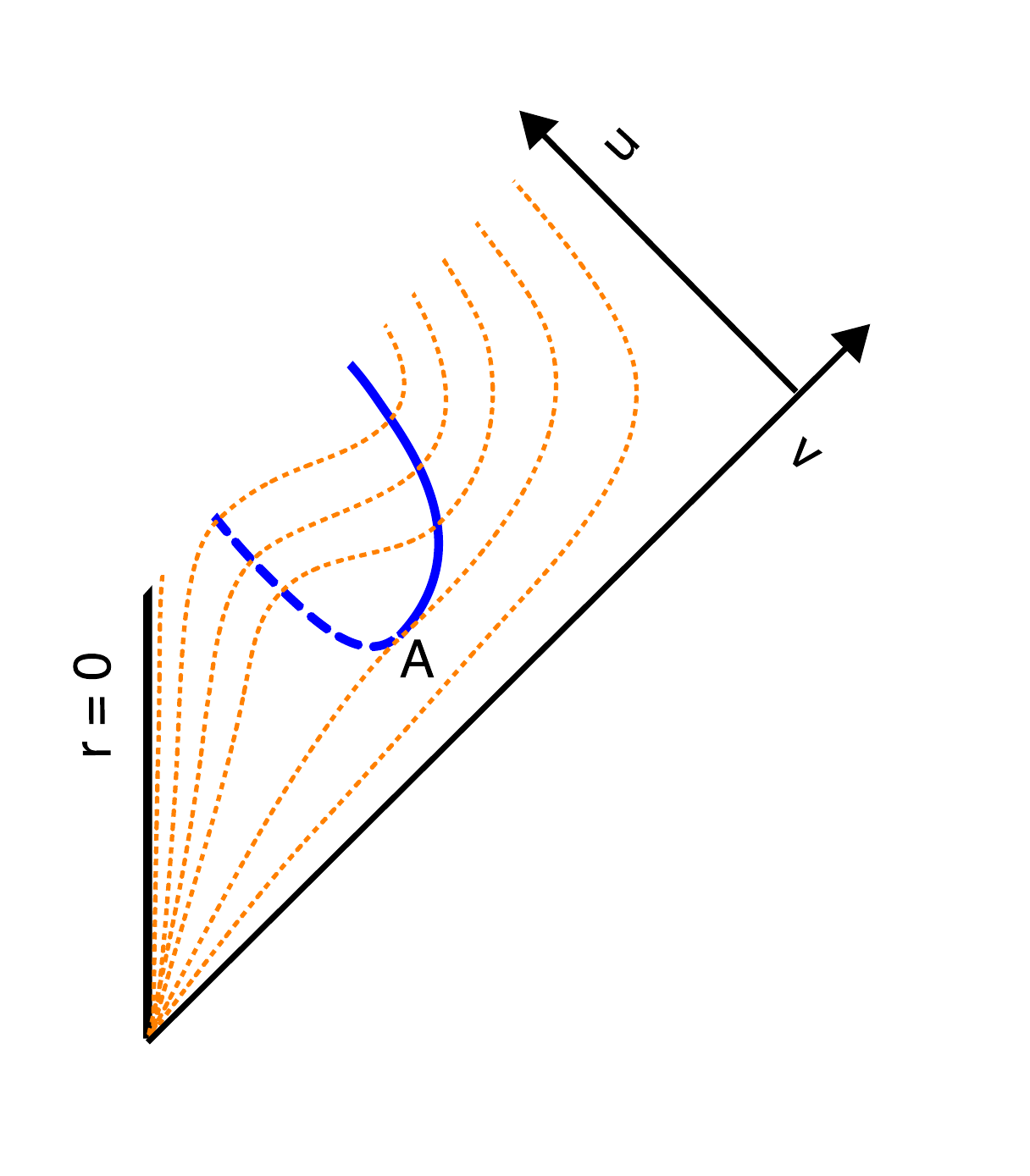}
\caption{\small
This is the Penrose diagram with a trapping horizon.
The solid and dashed curves (in blue) represent 
the trapping horizon in vacuum and that in matter,
respectively.
These two curves meet at {\sffamily A},
which is the point on the trapping horizon
with the lowest value of the $u$-coordinate.
The dotted curves (in orange) are constant $r$-curves,
whose tangents are light-like on the trapping horizon.
}
\label{trapping-horizon}
\vskip0em
\end{figure}

Let us assume that there is a trapped region
and thus a trapping horizon 
for the black hole under consideration.
The Penrose diagram with a trapping horizon 
in the absence of singularity
is schematically shown in Fig.\ref{trapping-horizon}.
We leave out the upper part of the Penrose diagram 
which may involve UV physics at least near the origin.
\footnote{
In the absence of singularity,
the trapping horizon should be a closed curve \cite{Hayward:2005gi}.
But it is possible that 
a regular geometric description is no longer valid 
at the origin.
%For a classical black hole,
%the trapping horizon is an open, space-like curve under the event horizon.
%This is possible only when there is singularity.
}

The trapping horizon is divided into two parts
by the point with the minimal value of the $u$-coordinate.
(It is marked by {\sffamily A} in Fig.\ref{trapping-horizon}.)
We will show below that 
the branch of the trapping horizon to the right of {\sffamily A}
is time-like and has $T_{vv} < 0$,
while the branch of the trapping horizon to the left of {\sffamily A}
is space-like and has $T_{vv} > 0$.
We shall refer to the former 
as the ``trapping horizon in vacuum''
and the latter as the ``trapping horizon in matter''.

We will focus on the trapping horizon in vacuum,
where
\be
\del_v^2 R > 0
\label{dv2r>0}
\ee
(unless there is degeneracy)
because $\del_v R$ is positive (negative) at slightly larger (smaller) $v$
(with $u$ fixed).

The outer trapping horizon defined in Ref.\cite{Hayward:2005gi},
includes both branches of the trapping horizon in Fig.\ref{trapping-horizon}.
On the outer trapping horizon,
one has $\del_u \del_v R < 0$.

Let the $u$-coordinate of the point {\sffamily A}
in Fig.\ref{trapping-horizon} be denoted $u_A$.
For $u > u_A$,
a constant $u$-curve intersects the trapping horizon at two points.
The trapping horizon in vacuum has the larger $v$-coordinate,
which will be denoted $v_0(u)$.
On the apparent horizon at $(u, v_0(u))$,
we have
\be
\frac{\del R}{\del v}(u, v_0(u)) = 0.
\ee
The expansions of $R(u, v)$ and $a(v)$ in powers of $(v-v_0(u))$ are
\begin{align}
R(u, v) &= R_0(u) + \frac{1}{2}R_2(u)(v-v_0(u))^2 + \cdots,
\label{R-exp}
\\
a(v) &= a(v_0(u)) + a'(v_0(u))(v-v_0(u)) + \cdots,
\label{a-exp}
\end{align}
in a small neighborhood of $v = v_0$.
We use primes and dots to indicate derivatives with respect to $v$ and $u$.

With the expansions \eqref{R-exp} and \eqref{a-exp},
we deduce from eq.\eqref{dRuvdv-null} that
%\begin{align}
%R_2(u) (v-v_0)
%&\simeq
%\frac{R_0 - a(v_0)}{2R_0}
%- \frac{a'(v_0)}{2R_0}(v-v_0),
%\end{align}
%which implies that
\begin{align}
R_0(u) = a(v_0(u)),
\quad \mbox{and} \quad
R_2(u) = - \frac{a'(v_0(u))}{2a(v_0(u))}.
\label{R-2}
\end{align}

According to eq.\eqref{dv2r>0}),
we have $R_2(u) > 0$,
so
\be
a'(v_0) < 0.
\label{ap<0}
\ee
As $T_{vv}$ is proportional to $a'(v)$ (see eq.\eqref{Tvv-a}),
the trapping horizon exists
only if the null energy condition is violated.

Plugging eq.\eqref{R-2}
into eq.\eqref{dRuvdu-null},
we find
\begin{align}
\dot{v}_0(u)
&\simeq
- \frac{C(u,v_0(u))}{2 a'(v_0(u))},
\end{align}
where $C(u, v)$ is given in eq.\eqref{C-sol}.
Due to eq.\eqref{ap<0},
we must have $\dot{v}_0 > 0$,
hence the trapping horizon in vacuum
is always time-like.

Incidentally,
the same analysis can be applied to 
the trapping horizon in matter,
and one would find $R_2 < 0$, $a' > 0$ and $\dot{v}_0 < 0$ instead.
The trapping horizon in matter is hence space-like,
and the null energy condition is satisfied.

Let us continue our study of the trapping horizon in vacuum.
Its tangent vector on the $(u, v)$-plane is
\begin{align}
%\xi = 
(\xi^u, \xi^v) \equiv
\frac{1}{\sqrt{C\dot{v}_0}} (1, \dot{v}_0)
\simeq \left(\frac{\sqrt{2|a'(v_0)|}}{C(u,v_0(u))}, \frac{1}{\sqrt{2|a'(v_0)|}}\right).
\end{align}
This is the unit time-like vector 
for an observer staying on top of the trapping horzon.
An orthonormal space-like vector in the radial direction is
\begin{align}
\chi = (\chi^u, \chi^v)
\simeq \left(\frac{\sqrt{2|a'(v_0)|}}{C(u,v_0(u))}, -\frac{1}{\sqrt{2|a'(v_0)|}}\right).
\end{align}

Around the trapping horizon where the ingoing Vaidya metric \eqref{ingoing-Vaidya-metric}
is a good approximation,
%$\langle T_{uu}\rangle = \langle T_{uv}\rangle = 0$,
the energy density on the trapping horizon in vacuum is
\begin{align}
{\cal E} \equiv \langle T_{vv}\rangle \xi^v \xi^v
\simeq 
- \frac{1}{2\ell_p^2 a^2(v_0)} < 0.
\label{E}
\end{align}

Notice that this result is independent of
both the details of the vacuum energy-momentum tensor
and that of the collapsing matter.
The only assumption in addition to spherical symmetry
is the existence of the trapping horizon,
and that the vacuum energy is dominated by $\langle T_{vv}\rangle$.
Notice also that ${\cal E}$ 
is proportional to $\ell_p^{-2}$,
hence it is of the same order 
as the (naive) mass density of the classical matter
(but with a minus sign). 
It diverges in the limit $\ell_p^2 \rightarrow 0$,
signaling the non-perturbative nature of this effect.
Furthermore,
${\cal E}$ is gauge-invariant
as the trapping horizon is independent of the choice of coordinate system
for spherically symmetric configurations.

In the classical limit $\hbar \rightarrow 0$,
the tangent vector $\xi$ of the trapping horizon is light-like,
so one might suspect that we find a large energy density 
simply because these observers are moving at a velocity 
close to the speed of light.
%%% 06/06+
%This is not completely correct,
%as the exact values of $\xi^u$ and $\xi^v$
%are actually of the same order of magnitude.
%Moreover,
Yet,
%%% 06/06-
while every local inertial frame is moving at nearly the speed of light
with respect to some other inertial frames,
the tangent vector $\xi$ of the trapping horizon
is the only natural choice of a local reference frame,
as the only gauge-invariant time-like vector there.

The amount of negative vacuum energy flowing into the trapping horizon in vacuum
(at the speed of light) per unit time is thus
the area of the apparent horizon times ${\cal E}$,
that is,
\begin{align}
P = 4\pi a^2(v_0) {\cal E} \simeq - \frac{2\pi}{\ell_p^2}.
\label{P}
\end{align}
This is equivalent to the negative mass of $- 10^{36}\;${\em kg} per second!

\subsection{Under Trapping Horizon}

\begin{figure}[h]
\vskip0em
\center
\includegraphics[scale=0.4,bb=0 30 450 260]{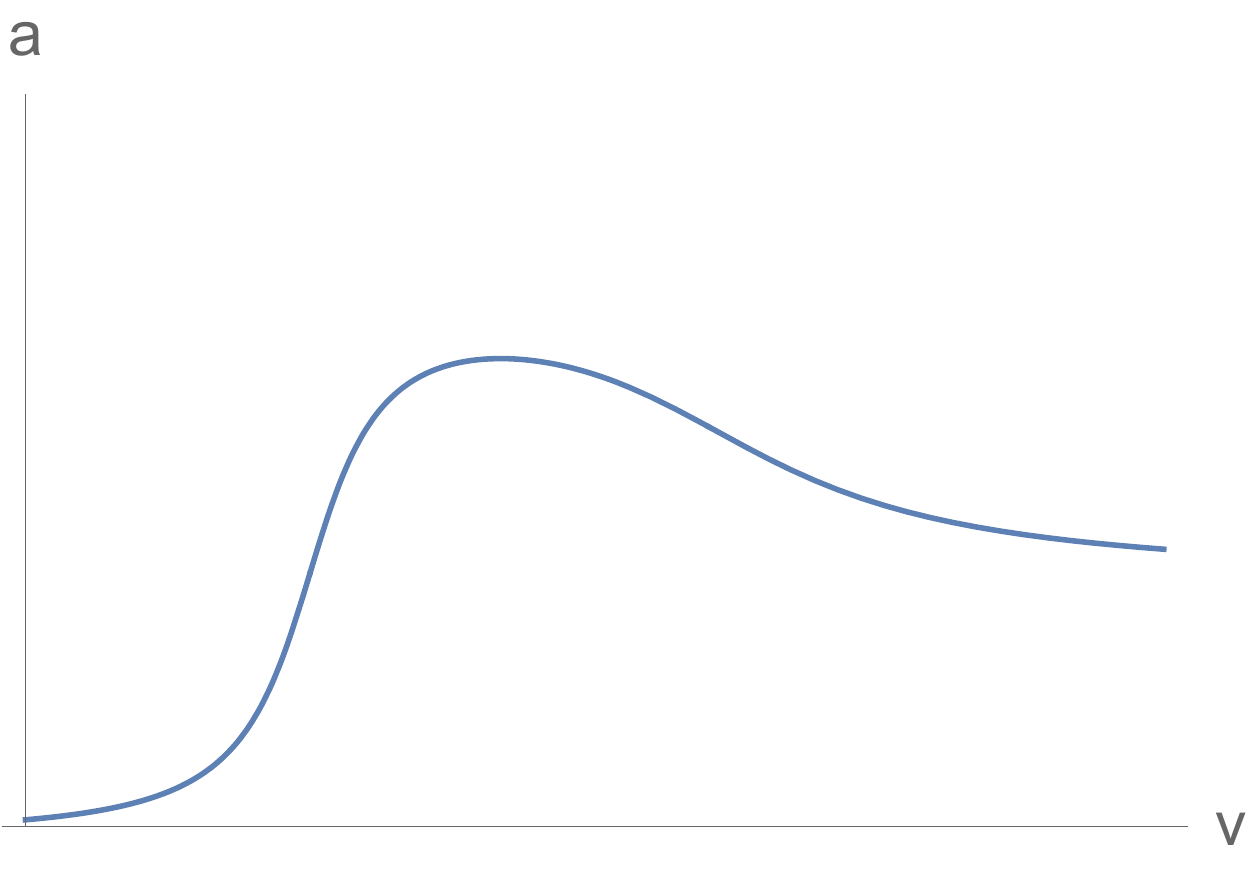}
\includegraphics[scale=0.4,bb=0 30 300 260]{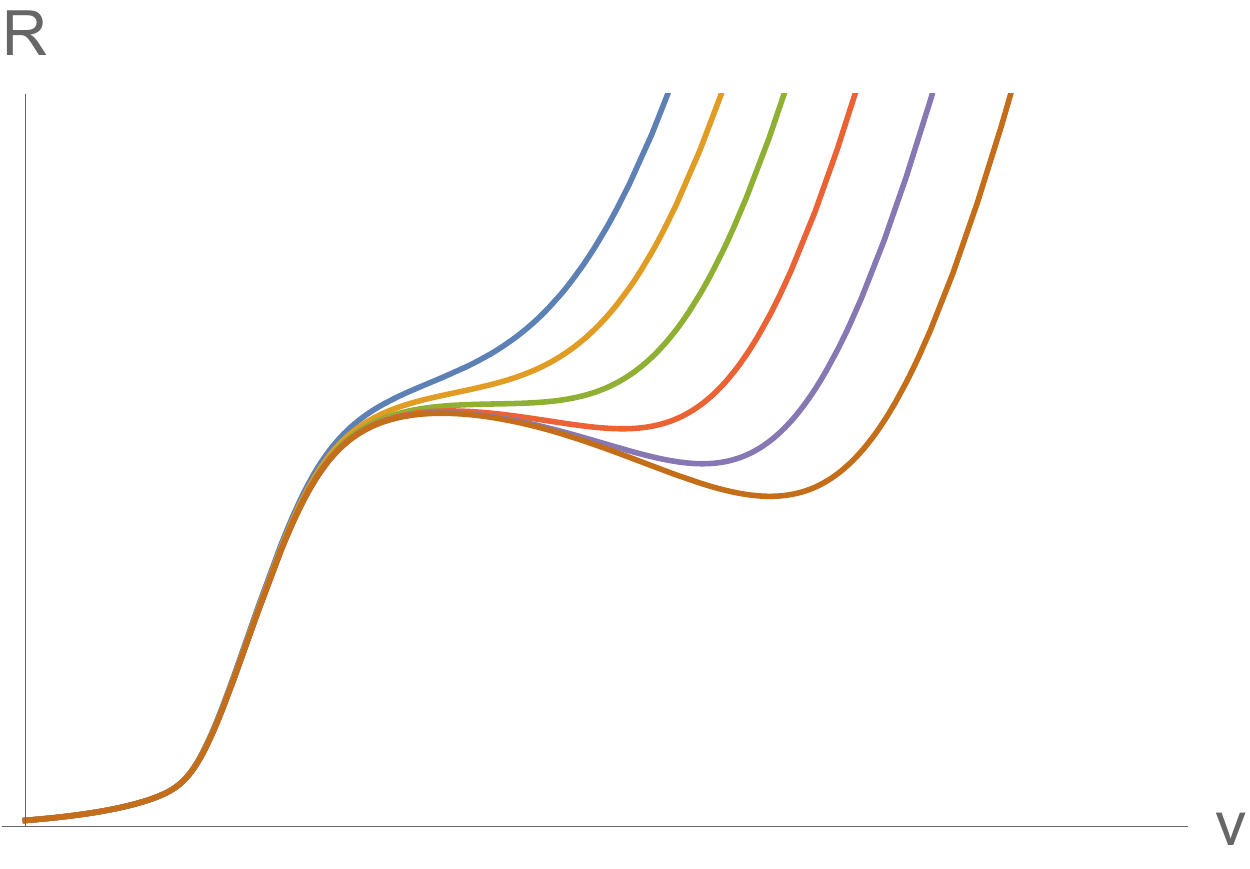}
\vskip1em
(a)\hskip17em (b)
\caption{\small
(a) Schematic $a-v$ diagram:
For small $v$,
$a$ increases with $v$ in the region occupied by collapsing matter.
At larger $v$,
$a$ decreases with $v$ due to 
the negative vacuum energy.
(The amount of negative energy in vacuum is exaggerated 
for demonstration.)
(b) Schematic $R-v$ plots over a sequence of $u$'s:
The apparent horizon is located at the local minimum of $R$,
where $\del R/\del v = 0$.
The areal radius of the apparent horizon shrinks as $u$ increases,
while the internal region (at small $v$)
is essentially frozen by the large red shift factor.
}
\label{Under-Horizon}
\vskip0em
\end{figure}

Although the calculation above is strictly speaking only valid 
around the trapping horizon,
the region under the trapping horizon in vacuum is essentially ``frozen''
by a huge red-shift factor,
i.e. it changes extremely slowly with $u$,
since $C(u, v)$ is extremely small.
We are therefore allowed to sketch the $R-v$ relation
under the trapping horizon in vacuum using our approximate description.
With a schematic profile of energy distribution $a(v)$
in Fig.\ref{Under-Horizon}(a),
a schematic behavior of $R$ is shown in Fig.\ref{Under-Horizon}(b),
by numerically solving eq.\eqref{dRuvdv-null}.
The function $a(v)$ goes to zero at $R=0$, 
and is an increasing function for small $v$ where the collapsing matter is. 
For larger $v$, 
outside the collapsing matter, 
$a(v)$ is a decreasing function because of the negative vacuum energy.

In the $R-v$ diagram in Fig.\ref{Under-Horizon}(b),
the value of $R$ at a large $v$ is given as a boundary condition
for each curve.
The local minimum of $R$ (the ``neck'') is where the apparent horizon is.
Due to eq.\eqref{dRuvdu-null},
a curve with a narrower ``neck'' corresponds to a larger value of $u$.
Fig.\ref{Under-Horizon}(b) is in agreement with the numerical simulation 
of the DFU model for a fully dynamical collapsing process \cite{Parentani:1994ij}.

Although the outgoing energy flux $\langle T_{uu} \rangle$
approximately vanishes at the trapping horizon in vacuum,
the total energy under the neck decreases over time
as the neck moves to the right in Fig.\ref{Under-Horizon}(b),
so that more negative energy is included under the neck.
The areal radius at the neck should be approximately
equal to the Schwarzschild radius.
It shrinks towards $0$,
and the geometry becomes reminiscent of the ``Wheeler's bag of gold''
\cite{Wheeler-Bag-Of-Gold}.
The low-energy effective theory breaks down 
before the areal radius at the neck is of Planck length.
We need a high energy theory to determine
whether or not the neck eventually shrinks to zero,
or whether the Wheeler's bag of gold detaches.

\section{Comments}
\label{conclusion}

For the conventional model of black holes,
the ingoing negative vacuum energy 
(with the power $P = - 2\pi/\ell_p^2$ \eqref{P})
is accumulated under the apparent horizon
so that eventually it cancels
the energy of the collapsed matter in the black hole.
With such a macroscopic negative energy in the conventional model,
one should not expect the holographic principle to hold.

This challenges the wide-spread belief that
there is no high-energy event around the black-hole horizon.
On the other hand,
we found a gauge-invariant quantity ${\cal E} \simeq -1/2\ell_p^2 a^2$ \eqref{E}
(the energy-density for an observer on top of the trapping horizon in vacuum)
that is inversely proportional to $\ell_p^2$. 
This implies that the quantum correction is at least comparable to the classical energy. 
While this may not immediately justify the need of a high-energy theory,
it opens such a possibility
and motivates further investigation,
e.g. the effect of related gauge-invariant terms
in the effective Lagrangian.

If the holographic principle should hold
for any consistent theory of quantum gravity,
one should rule out all models of vacuum energy-momentum tensor
in which a black hole loses energy 
mainly due to ingoing negative vacuum energy.
An alternative is to choose models in which
the vacuum energy-momentum tensor around a dense collapsing matter
is dominated by Hawking radiation.
In fact, 
the implication of this assumption about vacuum energy-momentum tensor
has been investigated in the KMY model \cite{Kawai:2013mda}.
(See also its follow-up works \cite{Kawai:2014afa}--\cite{Kawai:2017txu}.)
It was found that there would be no apparent horizon,
but there can be Planck-scale pressure at the surface of the collapsing matter
which signals the breakdown of low-energy effective theories.

It would be interesting to see more rigorously
how different assumptions about the vacuum energy-momentum tensor
is associated with the existence of trapping horizon
and the accumulation of macroscopic negative energy.
We leave this question for future works.

\section*{Acknowledgements}

The authors would like to thank 
Heng-Yu Chen,
Chong-Sun Chu,
Yu-tin Huang,
Hikaru Kawai,
Yutaka Matsuo,
Ioannis Papadimitriou,
Wen-Yu Wen,
and Piljin Yi
for discussions.
The work is supported in part by
the Ministry of Science and Technology, R.O.C.
(project no.\ 107-2119-M-002-031-MY3)
and by National Taiwan University
(project no.\ 105R8700-2).
The work of Y.M. (from April 2019) 
is supported in part by 
JSPS KAKENHI
Grants No.~JP17H06462.

% example of figure and figure in the margin 
% can be found at the end of the file.

%\end{CJK} 
\end{document}